\documentclass[a4paper,conference]{IEEEtran}

\ifCLASSINFOpdf
 
\else

\fi
\usepackage{fullpage,subfigure}

\usepackage{graphicx,amsmath,amsfonts,amscd,amssymb,bm,cite,epsfig,epsf,url,color}
\usepackage[small,bf]{caption}

\begin{document}
%

\newtheorem{theorem}{Theorem}[section]
\newtheorem{lemma}[theorem]{Lemma}
\newtheorem{corollary}[theorem]{Corollary}
\newtheorem{proposition}[theorem]{Proposition}
\newtheorem{definition}[theorem]{Definition}
\newtheorem{conjecture}[theorem]{Conjecture}
\newtheorem{remark}[subsection]{Remark}
\newtheorem{remarks}[subsection]{Remarks}
\newtheorem{example}[subsection]{Example}

\newcommand{\R}{\mathbb{R}}
\newcommand{\C}{\mathbb{C}}
\newcommand{\K}{\mathbb{K}}
\newcommand{\Z}{\mathbb{Z}}
\newcommand{\<}{\langle}
\renewcommand{\>}{\rangle}
\newcommand{\Var}{\textrm{Var}}
\newcommand{\goto}{\rightarrow}
\newcommand{\sgn}{\textrm{sgn}}
\renewcommand{\P}{\operatorname{\mathbb{P}}}
\newcommand{\E}{\operatorname{\mathbb{E}}}
\newcommand{\norm}[1]{{\left\lVert{#1}\right\rVert}}
\newcommand{\col}{\textrm{col}}

\newcommand{\TODO}[1]{{\bf TODO: #1}}
\renewcommand{\vec}[1]{{\boldsymbol{#1}}}
\newcommand{\e}{\mathrm{e}}
\renewcommand{\i}{\imath}

\newcommand{\ud}{\underline{\delta}}

\newcommand{\vct}[1]{\bm{#1}}
\newcommand{\mtx}[1]{\bm{#1}}

\newcommand{\bx}{\bm{x}}
\newcommand{\bX}{\bm{X}}
\newcommand{\bH}{\bm{H}}
\newcommand{\bM}{\bm{M}}
\newcommand{\bI}{\bm{I}}
\newcommand{\bU}{\bm{U}}
\newcommand{\bY}{\bm{Y}}
\newcommand{\bZ}{\bm{Z}}
\newcommand{\bb}{\bm{b}}
\newcommand{\by}{\bm{y}}
\newcommand{\bu}{\bm{u}}
\newcommand{\bv}{\bm{v}}
\newcommand{\be}{\bm{e}}
\newcommand{\bz}{\bm{z}}
\newcommand{\bzi}{\bm{z_i}}

\newcommand{\transp}{T}
\newcommand{\adj}{*}
\newcommand{\psinv}{\dagger}

\newcommand{\lspan}[1]{\operatorname{span}{#1}}

\newcommand{\range}{\operatorname{range}}
\newcommand{\colspan}{\operatorname{colspan}}

\newcommand{\rank}{\operatorname{rank}}

\newcommand{\diag}{\operatorname{diag}}
\newcommand{\tr}{\operatorname{Tr}}

\newcommand{\supp}[1]{\operatorname{supp}(#1)}

\newcommand{\smax}{\sigma_{\max}}
\newcommand{\smin}{\sigma_{\min}}

\newcommand{\restrict}[1]{\big\vert_{#1}}

\newcommand{\Id}{\text{\em I}}
\newcommand{\OpId}{\mathcal{I}}

\numberwithin{equation}{section}

\def \endprf{\hfill {\vrule height6pt width6pt depth0pt}\medskip}
\newenvironment{proof}{\noindent {\em Proof} }{\endprf\par}

\newcommand{\iprod}[2]{\left\langle #1 , #2 \right\rangle}

\newcommand{\red}{\textcolor{red}}

\newcommand{\ejc}[1]{\textcolor{red}{[EJC: #1]}}

\newcommand{\srf}{\text{SRF}}
\newcommand{\srfeq}{\text{\emph{SRF}}}
\newcommand{\I}{{\mathbb I}}
\newcommand{\Prob}[1]{\mathbb{P} \left\{ #1 \right\}}
\newcommand{\nil}{\varnothing}
\newcommand{\Cov}{\operatorname{Cov}}
\newcommand{\Proj}[2]{\mathcal{P}_{#2} ( #1 )}
\newcommand{\Projb}[2]{\mathcal{P}_{#2} \left( #1 \right)}
\newcommand{\Projn}[1]{\mathcal{P}_{#1}}
\newcommand{\vect}[1]{\text{vec}\left( #1 \right)}
\newcommand{\PTeps}[1]{\mathcal{P}_{T_{\epsilon}} \left( #1 \right)}
\newcommand{\PT}[1]{\mathcal{P}_{\mathcal{T}} \left( #1 \right)}
\newcommand{\PTd}[1]{\mathcal{P}_{\mathcal{T}_d} \left( #1 \right)}
\newcommand{\PTepsn}{\mathcal{P}_{T_{\epsilon}} }
\newcommand{\PTepsT}{\mathcal{P}_{T_{\epsilon}/T} }
\newcommand{\PTepsc}[1]{\mathcal{P}_{T_{\epsilon}^c} \left( #1 \right)}
\newcommand{\PTepscn}[1]{\mathcal{P}_{T_{\epsilon}^c} #1 }
\newcommand{\PTn}{P_{T} }
\newcommand{\PTdn}{\mathcal{P}_{\mathcal{T}_d} }
\newcommand{\PTp}[1]{\mathcal{P}_{\mathcal{T}^{\perp}} \left( #1 \right)}
\newcommand{\PTepsp}[1]{\mathcal{P}_{T_{\epsilon}^{\perp}} \left( #1 \right)}
\newcommand{\PTdp}[1]{\mathcal{P}_{\mathcal{T}_d^{\perp}} \left( #1 \right)}
\newcommand{\PTpn}{\mathcal{P}_{\mathcal{T}^{\perp}}}
\newcommand{\PTcn}{P_{T^{c}}}
\newcommand{\PTepspn}{\mathcal{P}_{T_{\epsilon}^{\perp}}}
\newcommand{\PTdpn}{\mathcal{P}_{\mathcal{T}_d^{\perp}}}
\newcommand{\PTt}[1]{\mathcal{P}_{\tilde{\mathcal{T}}} \left( #1 \right)}
\newcommand{\PTtn}{\mathcal{P}_{\tilde{\mathcal{T}}} }
\newcommand{\PTtp}[1]{\mathcal{P}_{\tilde{\mathcal{T}}^{\perp}} \left( #1 \right)}
\newcommand{\PTtpn}{\mathcal{P}_{\tilde{\mathcal{T}}^{\perp}}}
\newcommand{\PTnc}[1]{\mathcal{P}_{T} \left( #1 \right)}
\newcommand{\PAT}[1]{\mathcal{P}_{\mathcal{A}_T} \left( #1 \right)}
\newcommand{\PATn}{\mathcal{P}_{\mathcal{A}_T} }
\newcommand{\PATp}[1]{\mathcal{P}_{\mathcal{A}^{\perp}_T} \left( #1 \right)}
\newcommand{\PATpn}{\mathcal{P}_{\mathcal{A}^{\perp}_T}}
\newcommand{\PA}[1]{\mathcal{P}_{\mathcal{A}} \left( #1 \right)}
\newcommand{\PAn}{\mathcal{P}_{\mathcal{A}} }
\newcommand{\PAp}[1]{\mathcal{P}_{\mathcal{A}^{\perp}} \left( #1 \right)}
\newcommand{\PApn}{\mathcal{P}_{\mathcal{A}^{\perp}}}
\newcommand{\PB}[1]{\mathcal{P}_{\mathcal{B}} \left( #1 \right)}
\newcommand{\PBn}{\mathcal{P}_{\mathcal{B}} }
\newcommand{\PBp}[1]{\mathcal{P}_{\mathcal{B}^{\perp}} \left( #1 \right)}
\newcommand{\PBpn}{\mathcal{P}_{\mathcal{B}^{\perp}}}
\newcommand{\PTxp}[1]{\mathcal{P}_{\left(\mathcal{T} \cup X \right)^{\perp}} \left( #1 \right)}
\newcommand{\PTpx}[1]{\mathcal{P}_{\mathcal{T}^{\perp}\cap X} \left( #1 \right)}
\newcommand{\PTRp}[1]{\mathcal{P}_{\mathcal{T}_\mathcal{R}^{\perp}} \left( #1 \right)}
\newcommand{\PTpR}[1]{\mathcal{P}_{\mathcal{T}^{\perp}\cap \mathcal{RV}_2 } \left( #1 \right)}
\newcommand{\PTR}[1]{\mathcal{P}_{\mathcal{T}_\mathcal{R}} \left( #1 \right)}
\newcommand{\PU}[1]{\mathcal{P}_{\mathcal{U}} \left( #1 \right)}
\newcommand{\PUn}{\mathcal{P}_{\mathcal{U}}}
\newcommand{\PUtwo}[1]{\mathcal{P}_{\mathcal{U}_2} \left( #1 \right)}
\newcommand{\PUtwop}[1]{\mathcal{P}_{\mathcal{U}_2^{\perp}} \left( #1 \right)}
\newcommand{\PVv}[1]{\mathcal{P}_{\mathcal{V}\cup\tilde{v}} \left( #1 \right)}
\newcommand{\PV}[1]{\mathcal{P}_{\mathcal{V}} \left( #1 \right)}
\newcommand{\PVp}[1]{\mathcal{P}_{\mathcal{V}^{\perp}} \left( #1 \right)}
\newcommand{\PVu}[2]{\mathcal{P}_{\mathcal{V}_{#2}} \left( #1 \right)}
\newcommand{\PVup}[2]{\mathcal{P}_{\mathcal{V}_{#1}^{\perp}} \left( #2 \right)}
\newcommand{\PVone}[1]{\mathcal{P}_{\mathcal{V}_1} \left( #1 \right)}
\newcommand{\PVonep}[1]{\mathcal{P}_{\mathcal{V}_1^{\perp}} \left( #1 \right)}
\newcommand{\PVtwo}[1]{\mathcal{P}_{\mathcal{V}_2} \left( #1 \right)}
\newcommand{\PVtwop}[1]{\mathcal{P}_{\mathcal{V}_2^{\perp}} \left( #1 \right)}
\newcommand{\POmega}[1]{\mathcal{P}_{\Omega} \left( #1 \right)}
\newcommand{\POmegac}[1]{\mathcal{P}_{\Omega^c} \left( #1 \right)}
\newcommand{\POmegan}{\mathcal{P}_{\Omega}}
\newcommand{\POmegacn}{\mathcal{P}_{\Omega^c}}
\newcommand{\POmegak}[1]{\mathcal{P}_{\Omega_k} \left( #1 \right)}
\newcommand{\POmegakc}[1]{\mathcal{P}_{\Omega^c_k} \left( #1 \right)}
\newcommand{\POmegakn}{\mathcal{P}_{\Omega_k}}
\newcommand{\POmegakcn}{\mathcal{P}_{\Omega^c_k}}
\newcommand{\PWn}{\mathcal{P}_{W}}
\newcommand{\Rom}{\mathcal{R}_{\Omega}}
\newcommand{\Romh}{\hat{\mathcal{R}}_{\Omega}}
\newcommand{\Romk}[1]{\hat{\mathcal{R}}_{\Omega_{#1}}}
\newcommand{\T}{\mathcal{T}}
\newcommand{\Tk}{\mathcal{T}_{k}}
\newcommand{\Sk}{\mathcal{S}_{k}}
\newcommand{\nOne}[1]{\left|\left| #1 \right|\right| _{l_{1}}}
\newcommand{\normA}[1]{\left|\left| #1 \right|\right| _{\mathcal{A}}}
\newcommand{\normD}[1]{\left|\left| #1 \right|\right| _{\mathcal{D}}}
\newcommand{\normNuc}[1]{\left|\left| #1 \right|\right| _{\ast}}
\newcommand{\normOp}[1]{\left|\left| #1 \right|\right|}
\newcommand{\normF}[1]{\left|\left| #1 \right|\right| _{\mathcal{F}}}
\newcommand{\normInf}[1]{\left|\left| #1 \right|\right| _{L_\infty}}
\newcommand{\normTwo}[1]{\left|\left| #1 \right|\right| _{2}}
\newcommand{\normOne}[1]{\left|\left| #1 \right|\right| _{1}}
\newcommand{\normLOne}[1]{\left|\left| #1 \right|\right| _{L_1}}
\newcommand{\normLTwo}[1]{\left|\left| #1 \right|\right| _{\mathcal{L}_2}}
\newcommand{\normSharp}[1]{\left|\left| #1 \right|\right| _{\sharp}}
\newcommand{\normFlat}[1]{\left|\left| #1 \right|\right| _{\flat}}
\newcommand{\normArrow}[3]{\left|\left| #1 \right|\right| _{#2 \rightarrow #3}}
\newcommand{\normInfInf}[1]{\left|\left| #1 \right|\right| _{\infty}}
\newcommand{\normp}[1]{\left|\left| #1 \right|\right| _{p}}
\newcommand{\normTV}[1]{\left|\left| #1 \right|\right| _{\text{TV}}}
\newcommand{\normTVtwo}[1]{\left|\left| #1 \right|\right| _{\text{TV}\brac{2}}}
\newcommand{\TV}[1]{\text{TV}\left( #1 \right)}
\newcommand{\maxMat}[1]{\max_{a,b}\left| #1_{ab} \right|}
\newcommand{\maxMatP}[1]{\max_{a,b} \left| \frac{ #1_{a b}}{p_{a,b}} \right|}
\newcommand{\MUV}{K_{UV^{\ast}}}
\newcommand{\sign}[1]{\text{sign}\left( #1 \right) }
\newcommand{\signEq}[1]{\emph{\text{sign}}\left( #1 \right) }
\newcommand{\trace}[1]{\text{Trace}\left( #1 \right) }
\newcommand{\inv}[1]{\left( #1 \right)^{-1} }
\newcommand{\abs}[1]{\left| #1 \right|}
\newcommand{\keys}[1]{\left\{ #1 \right\}}
\newcommand{\sqbr}[1]{\left[ #1 \right]}
\newcommand{\brac}[1]{\left( #1 \right) }
\newcommand{\MAT}[1]{\begin{bmatrix} #1 \end{bmatrix}}
\newcommand{\sMAT}[1]{\left(\begin{smallmatrix} #1 \end{smallmatrix}\right)}
\newcommand{\sMATn}[1]{\begin{smallmatrix} #1 \end{smallmatrix}}
\newcommand{\PROD}[2]{\left \langle #1, #2\right \rangle}
\newcommand{\MATR}[3]{\left( \begin{array}{cc}  #1 & #2\\ \multicolumn{2}{c}{#3} \end{array} \right)}
\newcommand{\MATRn}[3]{ \begin{array}{cc}  #1 & #2\\ \multicolumn{2}{c}{#3} \end{array} }
\newcommand{\der}[2]{\frac{\text{d}#2}{\text{d}#1}}
\newcommand{\derTwo}[2]{\frac{\text{d}^2#2}{\text{d}#1^2}}
\newcommand{\ceil}[1]{\lceil #1 \rceil}
\newcommand{\Imag}[1]{\text{Im}\brac{ #1 }}
\newcommand{\Real}[1]{\text{Re}\brac{ #1 }}
\newcommand{\SN}{S_{\text{near}}}
\newcommand{\SF}{S_{\text{far}}}
\newcommand{\SFEq}{S_{\emph{{\text{far}}}}}
\newcommand{\SNEq}{S_{\emph{\text{near}}}}

\newcommand{\cA}{\mathcal{A}}

\newcommand{\optvalue}{2}
\newcommand{\optvaluetimestwo}{4}
\newcommand{\optvaluetwoD}{2.38}
\newcommand{\minm}{128}
\newcommand{\minmTwoD}{512}
\newcommand{\tOne}{1.0699}
\newcommand{\tTwo}{0.6814}
\newcommand{\tC}{0.1649}
\newcommand{\tp}{t_{+}}
\newcommand{\tm}{t_{-}}
\newcommand{\tA}{0.7559}
\newcommand{\tx}{0.4269}
\newcommand{\ty}{0.8829}
\newcommand{\rOne}{0.2447}
\newcommand{\rTwo}{0.84}
\newcommand{\fc}{f_c}
\newcommand{\Deltamin}{\Delta_{\text{min}}}
\newcommand{\Deltaminth}{\Delta_{\text{\em min}}}
\newcommand{\jp}{20}
\newcommand{\Ktwo}{K^{\text{2D}}}
\newcommand{\Ker}{K}
\newcommand{\low}{c}
\newcommand{\high}{\text{hi}}
\newcommand{\Klo}{K_{\low}}
\newcommand{\Khi}{K_{\high}}
\newcommand{\KhiEq}{K_{\emph{\high}}}
\newcommand{\flo}{f_{\low}}
\newcommand{\floEq}{f_{\emph{\low}}}
\newcommand{\fhi}{f_{\high}}
\newcommand{\lambdalo}{\lambda_{\low}}
\newcommand{\lambdaloEq}{\lambda_{\emph{\low}}}
\newcommand{\lambdahi}{\lambda_{\high}}
\newcommand{\lambdahiEq}{\lambda_{\emph{\high}}}
\newcommand{\Qlo}{Q_{\low}}
\newcommand{\QloEq}{Q_{\emph{\low}}}
\newcommand{\Qhi}{Q_{\high}}
\newcommand{\Flo}{F_{\low}}
\newcommand{\FloEq}{F_{\emph{\low}}}
\newcommand{\dEq}{\emph{\text{d}}}
\newcommand{\xest}{x_{\text{est}}}
\newcommand{\xestEq}{x_{\emph{\text{est}}}}
\newcommand{\uest}{u_{\text{est}}}
\newcommand{\uestEq}{u_{\emph{\text{est}}}}
\newcommand{\normTVEq}[1]{\left|\left| #1 \right|\right| _{\emph{\text{TV}}}}

\newcommand{\cfg}[1]{\textcolor{blue}{#1}}

\title{Support detection in super-resolution}

\author{\IEEEauthorblockN{Carlos Fernandez-Granda}
\IEEEauthorblockA{Department of Electrical Engineering, Stanford University\\
Stanford, CA, USA\\
Email: cfgranda@stanford.edu}}

\maketitle

\begin{abstract}
We study the problem of super-resolving a superposition of point sources from noisy low-pass data with a cut-off frequency $\fc$. Solving a tractable convex program is shown to locate the elements of the support with high precision as long as they are separated by $2/\fc$ and the noise level is small with respect to the amplitude of the signal.
\end{abstract}

\IEEEpeerreviewmaketitle
\section{Introduction}
The problem of super-resolution is of great importance in applications where the measuring process imposes a physical limit on the resolution of the available measurements. It is often the case that the signal of interest is well modeled as a superposition of point sources. Motivated by this, we consider a signal 
\begin{equation}
  \label{eq:model}
  x = \sum_j a_j \delta_{t_j}, 
\end{equation}
consisting of a train of Dirac measures with complex amplitudes $a_j$ located at different locations $\{t_j\}$ in the unit interval. Our aim is to estimate $x$ from the lower end of its spectrum in the form of $n = 2\fc + 1$ Fourier series coefficients ($\fc$ is an integer) perturbed by noise,
\begin{align}
  y(k) = \int_0^1 e^{-i2\pi kt} x(\text{d}t) +z(k) \notag \\
   = \sum_j a_j e^{-i2\pi k t_j} + z(k),  \label{eq:fourier}
\end{align}
for $k \in \Z$, $\abs{k}\leq \fc$. To ease notation, we write \eqref{eq:fourier} as $y =
\mathcal{F}_{n} \, x +z$. We model the perturbation $z \in \C^n$ as having bounded $\ell_2$ norm,
\begin{equation}
  \label{eq:error}
\normTwo{z } \le \delta.
\end{equation}
The noise is otherwise arbitrary and can be adversarial. 

Even if the signal $x$ is very sparse, without further conditions to ensure that the support of $x$ is not too clustered the super-resolution problem is hopelessly ill-posed. This can be checked numerically, but also formalized thanks to the seminal work of Slepian \cite{slepian} on discrete prolate spheroidal sequences (see Section~3.2 of \cite{superres}). To avoid such extreme ill-posedness, we impose a lower bound on the minimum separation between the elements of the support of the signal.
\emph{\begin{definition}[Minimum separation] Let $\mathbb{T}$ be the circle
  obtained by identifying the endpoints on $[0,1]$. For a family of points $T \subset
  \mathbb{T}$, the minimum separation is the closest
  distance between any two elements of $T$,
  \begin{equation}
    \label{eq:min-distance-def}
    \Delta(T) = \inf_{(t, t') \in T \, : \, t \neq t'} \, \, |t - t'|. 
  \end{equation}
\end{definition}}
To recover $x$ we propose minimizing the total
variation of the estimate, a continuous analog to the $\ell_1$ norm for discrete signals (see Appendix A in~\cite{superres} for a rigorous definition), subject to data constraints:
\begin{align}
\label{TVproblem_relaxed}
\min_{\tilde x} \,  \normTV{\tilde x} 
\quad \text{subject to} \quad \normTwo{\mathcal{F}_{n} \tilde x - y} \leq \delta, 
\end{align}
where the minimization is carried out over the set of all finite
complex measures $\tilde x$ supported on $[0,1]$. For details on how to solve~\eqref{TVproblem_relaxed} using semidefinite programming see~\cite{superres,superres_noise}.
 
Previous work established that if
\begin{align}
\Delta(T) \geq \frac{2}{\fc}:= 2 \lambdaloEq \label{eq:min-distance}
\end{align}
TV-norm minimization achieves exact recovery in a noiseless setting~\cite{superres}. Additionally, \cite{superres_noise} characterized the reconstruction error for noisy measurements as the target resolution increases. In this work we study support detection using this method. If the original signal contains a spike of a certain amplitude we ask: \emph{How accurately can we recover the position of the spike? How does the accuracy depend on the noise level, the amplitude of the spike and the amplitude of the signal at other locations?} These questions are not addressed by previous work and answering them requires non-trivial modifications to the arguments in~\cite{superres_noise} and \cite{superres}. Our main result establishes that convex programming is in fact a powerful method for support detection in super-resolution.
\emph{\begin{theorem}
\label{theorem:support}
Consider the noise model~\eqref{eq:error} and assume the support $T$ satisfies the minimum-separation condition~\eqref{eq:min-distance}. The solution to problem~\eqref{TVproblem_relaxed}\footnote{This solution can be shown to be an atomic measure with discrete support under very general conditions.}
\begin{align*}
\hat{x} = \sum_{\hat{t}_k\in \widehat{T}} \hat{a}_k \delta_{\hat{t}_k}
\end{align*}
with support $\widehat{T}$ obeys the properties
\begin{align*}
&\text{(i)} \emph{:} \; \Big| a_j - \sum_{\keys{\hat{t}_l \in \widehat{T}: \, \abs{\hat{t}_l-t_j} \leq c \lambdaloEq }} \hat{a}_l \Big| \leq C_1 \delta  \quad \forall t_j \in T, \\
 & \text{(ii)}\emph{:} \; \sum_{\keys{\hat{t}_l \in \widehat{T}, \, t_j \in T: \, \abs{\hat{t}_l-t_j} \leq c \lambdaloEq }} \abs{\hat{a}_l} \brac{\hat{t}_l-t_j}^2  \leq C_2 \lambdaloEq^2 \delta ,\\ 
 & \text{(iii)} \emph{:} \; \sum_{\keys{\hat{t}_l \in \widehat{T}: \, \abs{\hat{t}_l-t_j}>c \lambdaloEq \, \forall t_j \in T} } \abs{\hat{a}_l}  \leq C_3 \delta,
\end{align*}
where $C_1$, $C_2$ and $C_3$ are positive numerical constants and $c=0.1649$. 
\end{theorem}}
Properties (i) and (ii) show that the estimate clusters tightly around each element of the signal, whereas (iii) ensures that any spurious spikes detected by the algorithm have small amplitude. These bounds are essentially optimal for the case of adversarial noise, which can be highly concentrated. An intriguing consequence of our result is a bound on the support-detection error for a single spike that does not depend on the value of the signal at other locations. 
\emph{\begin{corollary}
Under the conditions of Theorem~\ref{theorem:support}, for any element $t_i$ in the support of $x$ such that $a_i>C_1 \delta$ there exists an element $\hat{t}_i$ in the support of the estimate $\hat{x}$ satisfying
\begin{align*}
\abs{t_i-\hat{t}_i} \leq \sqrt{ \frac{ C_2 \delta}{\abs{a_i}-C_1\delta}}\lambda_c.
\end{align*}
\end{corollary}}
Despite the aliasing effect of the low-pass filter applied to the signal, the bound on the support-detection error \emph{only depends on the amplitude of the corresponding spike} (and the noise level). This does not follow from previous analysis. In particular, the bound on the weighted $\mathcal{L}_1$ norm of the error derived in~\cite{superres_noise} does not allow to derive such local guarantees. A recent paper bounds the support-detection error of a related convex program in the presence of stochastic noise, but the bound depends on the amplitude of the solution rather than on the amplitude of the original spike~\cite{spikes_azais}. As we explain below, obtaining detection guarantees that only depend on the amplitude of the spike of interest is made possible by the existence of a certain low-frequency polynomial, constructed in Lemma~2.2. This is the main technical contribution of the paper.  

\section{Proof of Theorem~\ref{theorem:support}}
We begin with an intermediate result proved in Section~\ref{proof:support}.
\emph{\begin{lemma}
\label{lemma:support}
Under the assumptions of Theorem~\ref{theorem:support}
 \begin{align*}
\sum_{ \hat{t}_k \in \hat{T}} \abs{\hat{a}_k} \min \keys{ C_a,\frac{C_b \, d\brac{\hat{t}_k,T}}{\lambdalo^2} }  & \leq 2 \delta,
\end{align*}
where $C_a$ and $C_b$ are positive numerical constants and
 \begin{align*}
d\brac{t,T} := \min_{t_i \in T}\brac{t-t_i}^2.
\end{align*}
\end{lemma}}
Properties (ii) and (iii) are direct corollaries of Lemma~\eqref{lemma:support}. To establish property (i) we need the following key lemma, proved in Section~\ref{section:q_tau}.
\emph{\begin{lemma}
\label{lemma:q_tau}
Suppose $T$ obeys condition~\eqref{eq:min-distance} and $\fc \geq 10$. Then for any $t_j \in T$ there exists a
low-pass polynomial 
\begin{align*}
q_{t_j}(t) = \sum_{k = -\floEq}^{\floEq} b_k e^{i2\pi k t},
\end{align*}
$b \in \C^{n}$, such that $|q_{t_j}(t)|  < 1$ for all $t \neq t_j$ and 
\begin{align}
& q_{t_j}(t_j)  = 1, \qquad \qquad q_{t_j}(t_l)  = 0 \quad  t_l \in T\setminus \keys{t_j}, \notag \\
&|1-q_{t_j}(t)|  \leq \frac{C'_{1}  \brac{t-t_j}^2}{\lambdaloEq^2}   \quad \text{for } \;   \abs{t-t_j}\leq  c \lambdaloEq, \label{qtau_2}\\
&|q_{t_j}(t)|  \leq \frac{C'_{1}  \brac{t-t_l}^2}{\lambdaloEq^2}   \quad \text{for } \;   t_l \in T\setminus \keys{t_j}, \; \abs{t-t_l}\leq  c \lambdaloEq,  \label{qtau_3}\\
& |q_{t_j}(t)|  < C'_{2}   \quad \text{if } \;    \abs{t-t_l}>  c \lambdaloEq \; \forall \; t_l \in T, \label{qtau_4}
\end{align}
where $0<c^2 C'_{2} \leq C'_{1}<1$.
\end{lemma}}
The polynomial $q_{t_j}$ provided by this lemma is designed to satisfy $\int_{\mathbb{T}}q_{t_j}(t)x(\text{d}t)=a_j$ and vanish on the rest of the support of the signal. This allows to decouple the estimation error at $t_j$ from the amplitude of the rest of the spikes. Since $x$ and $\hat{x}$ are feasible for~\eqref{TVproblem_relaxed}, we can apply Parseval's Theorem and the Cauchy-Schwarz inequality to obtain
\begin{align}
 & \abs{\int_{\mathbb{T}}q_{t_j}(t)x(\text{d}t)-\int_{\mathbb{T}}q_{t_j}(t)\hat{x}(\text{d}t)} \notag\\ 
& \qquad= \Big|\sum_{k = -\floEq}^{\floEq} b_k \mathcal{F}_n(x - \hat{x})_k \Big|  \notag \\
& \qquad \leq \normLTwo{q_{t_j}} \normTwo{\mathcal{F}_n(x - \hat{x})} \leq 2 \delta, \label{eq:2delta}
\end{align}
where we have used that the absolute value and consequently the $\mathcal{L}_2$ norm of $q_{t_j}$ is bounded by one. In addition, by Lemmas~\ref{lemma:q_tau} and~\ref{lemma:support} we have
\begin{align}
&\Big| \sum_{\keys{k: \, \abs{\hat{t}_k-t_j}>c \lambdalo }} \hat{a}_k q_{t_j}(\hat{t}_k) + \sum_{\keys{k: \, \abs{\hat{t}_k-t_j} \leq c \lambdalo }} \hat{a}_k \brac{ q_{t_j}(\hat{t}_k)-1} \Big| \notag \\
& \leq \sum_{\keys{k: \, \abs{\hat{t}_k-t_j}>c \lambdalo }} \abs{\hat{a}_k} \abs{q_{t_j}(\hat{t}_k)} \notag
\\ &\quad 
+\sum_{\keys{k: \, \abs{\hat{t}_k-t_j} \leq c \lambdalo }} \abs{\hat{a}_k} \abs{1-q_{t_j}(\hat{t}_k)} \notag \\ 
&\leq  \sum_{\hat{t}_k \in \hat{T}} \abs{\hat{a}_k} \min \keys{ C'_{2}, \frac{C'_{1} d\brac{\hat{t}_k,T}}{\lambdalo^2} } \leq C  \delta,\label{eq:Cdelta} 
\end{align}
for a positive numerical constant $C$. Finally, Lemma~\ref{lemma:q_tau}, the triangle inequality, \eqref{eq:2delta} and~\eqref{eq:Cdelta} yield 
\begin{align*}
& \Big| a_j - \sum_{\keys{k: \, \abs{\hat{t}_k-t_j} \leq c \lambdaloEq }} \hat{a}_k \Big| \\
& = \Bigg| \int_{\mathbb{T}}q_{t_j}(t)x(\text{d}t)-\int_{\mathbb{T}}q_{t_j}(t)\hat{x}(\text{d}t) \\
& \quad + \sum_{\keys{k: \, \abs{\hat{t}_k-t_j} > c \lambdalo }} \hat{a}_k q_{t_j}(\hat{t}_k) \\
& \quad + \sum_{\keys{k: \, \abs{\hat{t}_k-t_j} \leq c \lambdalo }} \hat{a}_k \brac{ q_{t_j}(\hat{t}_k)-1} \Bigg|  \leq C'  \delta.
\end{align*} 
for a positive numerical constant $C'$.
\subsection{Proof of Lemma~\ref{lemma:support}}
\label{proof:support}
The proof relies on a low-pass
polynomial provided by Proposition~2.1 and Lemma~2.5 in~\cite{superres}.
\emph{\begin{lemma}
\label{lemma:dualpol}
Let $T$ obey \eqref{eq:min-distance}. For any
$v \in \C^{|T|}$ such that $\abs{v_j}=1$ for all entries $v_j$ there exists a
low-pass polynomial $q(t) = \sum_{k = -\floEq}^{\floEq} d_k e^{i2\pi k t} $, $d \in \C^{n}$, satisfying
\begin{align*}
q(t_j) & = v_j, \qquad t_j \in T, \\
|q(t)| & < 1-C_a ,  \qquad  \abs{t-t_j}>  c \lambdaloEq \quad \forall t_j \in T, \\
|q(t)| & \leq 1-\frac{C_b  \brac{t-t_j}^2}{\lambdaloEq^2} ,  \qquad  \abs{t-t_j}\leq  c \lambdaloEq,\,t_j \in T, 
\end{align*}
with $0<c^2 C_b \leq C_a<1$.
\end{lemma}}
We set $v_j = \overline{a_j}/\abs{a_j}$. The lemma implies
 \begin{align}
& \int_{\mathbb{T}} q(t) \hat{x}\brac{\text{d}t} \leq \sum_{k} \abs{\hat{a}_k}\abs{q(\hat{t}_k)} \notag\\
\quad & \leq \sum_{k} \brac{1-\min \keys{ C_a,\frac{C_b \, d\brac{\hat{t}_k,T}}{\lambdalo^2} } } \abs{\hat{a}_k}.\label{eq:prod1}
\end{align}
The same argument used to prove~\eqref{eq:2delta} yields
\begin{align*}
 \abs{\int_{\mathbb{T}}q(t)\hat{x}(\text{d}t)-\int_{\mathbb{T}}q(t)x(\text{d}t)} \leq 2 \delta. 
\end{align*}
Now, taking into account that $\int_{\mathbb{T}} q(t) x\brac{\text{d}t}=\normTV{x}$ by construction and $\normTV{\hat{x}} \leq \normTV{x}$, we have
\begin{align}
& \int_{\mathbb{T}} q(t) \hat{x}\brac{\text{d}t} \notag\\ 
&\quad = \int_{\mathbb{T}} q(t) x\brac{\text{d}t}  +\int_{\mathbb{T}}q(t)\hat{x}(\text{d}t)-\int_{\mathbb{T}}q(t)x(\text{d}t) \notag \\
&\quad  \geq \normTV{x}- 2\delta   \geq \normTV{\hat{x}}- 2\delta   = \sum_{k}\abs{\hat{a}_k}- 2\delta .\notag
\end{align}
Combining this with~\eqref{eq:prod1} completes the proof.

\subsection{Proof of Lemma~\ref{lemma:q_tau}}
\label{section:q_tau}
We use a low-frequency kernel and its derivative to construct the desired polynomial exploiting the assumption that the support
satisfies the minimum-separation condition~\eqref{eq:min-distance}. More precisely, we set
\begin{equation}
  q_{t_j}(t)  = \sum_{t_k \in T} \alpha_k \Ker(t-t_k) + \beta_k \Ker^{\brac{1}}(t-t_k), 
  \label{def:q_tau}
\end{equation}
where $\alpha, \beta \in \C^{\abs{T}}$ are coefficient vectors,
\begin{equation} 
  \Ker(t) = \left[\frac{\sin \brac{\brac{\frac{\flo}{2}+1} \pi
        t}}{\brac{\frac{\flo}{2}+1}\sin \brac{\pi t}}\right]^4, \quad
  t \in \mathbb{T} \setminus\{0\}, 
\label{def:kernel}
\end{equation}
and $\Ker(0) = 1$; here, $\Ker^{\brac{\ell}}$ is the $\ell$th
derivative of $\Ker$. Note that $\Ker$, $\Ker^{\brac{1}}$ and,
consequently, $q_{t_j}$ are trigonometric polynomials of the required degree.

We impose $q_{t_j}(t_j) = 1$ and
\begin{align*}
q_{t_j}(t_l)  = 0,\quad t_l \in T/ \keys{t_j}, \qquad q_{t_j}'(t_k)  = 0,\quad t_k \in T.
\end{align*}
We express these constraints in
matrix form. Let $e_{t_j} \in \R^{|T|}$ denote the one-sparse vector with one nonzero entry at the position corresponding to $t_j$. Then,
\[
\begin{bmatrix} D_0 & D_1\\ D_1 & D_2 \end{bmatrix} \begin{bmatrix} \alpha \\
  \beta \end{bmatrix} =\begin{bmatrix} e_{t_j}\\  0 \end{bmatrix}, \quad \text{where}
\]
\begin{align*}
\brac{D_0}_{lk} &= \Ker\brac{t_l - t_k}, \quad \brac{D_1}_{lk} =
\Ker^{\brac{1}}\brac{t_l - t_k}, \\
\brac{D_2}_{lk}& = \Ker^{\brac{2}}\brac{t_l - t_k},
\end{align*}
and $l$ and $k$ range from $1$ to $\abs{T}$. It is shown in Section 2.3.1 of \cite{superres} that under the minimum-separation condition this system is invertible. As a result $\alpha$ and $\beta$ are well defined and $q_{t_j}$ satisfies $q_{t_j}(t_j)=1$ and $q_{t_j}(t_l)  = 0$ for $t_l \in T/ \keys{t_j}$. The coefficient vectors can be expressed as
\[
  \begin{bmatrix} \alpha \\
    \beta \end{bmatrix} =\begin{bmatrix} \Id \\
    -D_2^{-1}D_1 \end{bmatrix} S^{-1} e_{t_j}, \quad S :=
  D_0-D_1D_2^{-1}D_1,
\]
where $S$ is the Schur complement. Let $\|M \|_\infty$ denote the usual infinity norm of a matrix $M$ defined as
$\|M\|_\infty = \max_{\|x\|_\infty = 1} \|Mx\|_\infty = \max_l \sum_k
|M_{lk}|$. We borrow some results from Section~2.3.1 in~\cite{superres},
\begin{align}
\normInfInf{\Id - S} & \leq 8.747 \, 10^{-3}  , \notag \\
\normInfInf{S^{-1}} & \leq  1+ 8.824 \, 10^{-3} , \notag \\
\normInfInf{I-S^{-1}} & \leq  \normInfInf{S^{-1}}\normInfInf{S-I} \leq 8.825 \, 10^{-3}, \notag \\
\normInfInf{\alpha- e_{t_j}} & \leq \normInfInf{I-S^{-1}}\normInfInf{ e_{t_j}} \notag \\
& \leq  8.825 \, 10^{-3},\label{eq:boundAlpha}\\
\normInfInf{\beta} & \leq 3.294 \, 10^{-2} \lambdalo. \label{eq:boundBeta}
\end{align}
Lemma~2.6 in~\cite{superres} allows to obtain 
\begin{align*}
K\brac{t} & \leq \frac{1}{\brac{\fc t}^4} \leq 0.333, \quad
K'\brac{t}   \leq \frac{4\pi}{\fc^3 t^4} \leq 4.18\, \fc, 
\end{align*}
for $\abs{t}>c\lambdalo$ as long $\fc \geq 10$. By the same lemma, if we set the minimum separation $\Deltamin$ to $2/f_c$
\begin{align*}
&\sum_{t_k \in T\setminus \{t_a,t_b\}} \abs{\Ker\brac{t-t_k}}\\
& \leq \sum_{l=0}^{\infty} \frac{1}{\brac{\fc \Deltamin (\frac{1}{2}+l)}^4} +\sum_{l=0}^{\infty} \frac{1}{\brac{\fc \Deltamin l}^4} \leq 1.083,\\
&\sum_{t_k \in T\setminus \{t_a,t_b\}} \abs{\Ker^{\brac{1}}\brac{t-t_k}} \\
& \leq \sum_{l=0}^{\infty} \frac{4\pi}{\fc^3\brac{ \Deltamin (\frac{1}{2}+l)}^4} +\sum_{j=0}^{\infty} \frac{4\pi}{\fc^3\brac{ \Deltamin l}^4} \leq 1.75 \, \fc, 
\end{align*}
where $t_a$ and $t_b$ are the two spikes nearest to $t$. Let $t_i$ be the element of $T/\keys{t_j}$ that is nearest to $t$. Combining these inequalities with~\eqref{eq:boundAlpha} and~\eqref{eq:boundBeta} proves that
\begin{align*}
 |q_{t_j}(t)| & =  \Big| \sum_{t_k \in T}\alpha_k \Ker\brac{t-t_k} + \sum_{t_k \in T} \beta_k \Ker^{\brac{1}}\brac{t-t_k}\Big|\notag\\
& \leq  \abs{\Ker\brac{t-t_j}} +\normInfInf{\alpha- e_{t_j}}\Big(\abs{\Ker\brac{t-t_j}}  \\
& \quad +\abs{\Ker\brac{t-t_i}} +  \sum_{t_k \in T\setminus \{t_i,t_j\}} \abs{\Ker\brac{t-t_k}} \Big)\\ 
& \quad+ \normInfInf{\beta}\Big(\abs{\Ker^{\brac{1}}\brac{t-t_j}}+\abs{\Ker^{\brac{1}}\brac{t-t_i}}\\
& \quad+  \sum_{t_k \in T\setminus \{t_i,t_j\}} \abs{\Ker^{\brac{1}}\brac{t-t_k}}\Big) \leq 0.69, 
\end{align*}
if $\abs{t-t_k}>  c \lambdaloEq$ for all $t_k \in T$ so that \eqref{qtau_4} holds. The proof is completed by two lemmas which prove
\eqref{qtau_2} and \eqref{qtau_3} and $|q_{t_j}(t)|  < 1$ for any $t$. They rely on the following bounds borrowed from equation~(2.25) in Section~2 of~\cite{superres},
\begin{equation}
  \label{eq:boundK}
\begin{alignedat}{2}
K\brac{t} & \geq
0.9539, 
& \qquad K^{\brac{2}}\brac{t}  & \leq -2.923 \, \fc^2,  \\
\abs{K^{\brac{1}}\brac{t}} & \leq 0.5595 \, \fc, & \qquad \abs{K^{\brac{2}}\brac{t}}& \leq 3.393 \, \fc^2,\\
  \abs{K^{\brac{3}}\brac{t}} & \leq 5.697 \,
\fc^3, 
\end{alignedat}
\end{equation}
and on the fact that, due to Lemma 2.7 in \cite{superres}, for any $t_0
\in T$ and $t \in \mathbb{T}$ obeying $\abs{t-t_0}\leq c
\lambdalo$,
\begin{align}
  \sum_{t_k \in T\setminus \{t_0\}} \abs{\Ker^{\brac{2}}\brac{t-t_k}}
  \leq 1.06 \, \flo^{2} \label{eq:boundsum_der2}\\
\sum_{t_k \in T\setminus \{t_0\}} \abs{\Ker^{\brac{3}}\brac{t-t_k}}
  \leq 18.6 \, \flo^{3} \label{eq:boundsum_der3}.
\end{align}
\emph{\begin{lemma}
\label{lemma:qtau_2}
For any $t $ such that $\abs{t-t_j} \leq
c \lambdaloEq$,  
\begin{align*}
1-4.07\brac{t-t_j}^2\flo^2 \leq q_{t_j}(t) \leq 1-2.30\brac{t-t_j}^2\flo^2.
\end{align*}
\end{lemma}}
\begin{proof}
We assume without loss of generality that $t_j = 0$. By symmetry, it suffices to show the claim for $t \in (0,c \,\lambdalo]$. 
By \eqref{eq:boundAlpha}, \eqref{eq:boundBeta}, \eqref{eq:boundK}, \eqref{eq:boundsum_der2} and \eqref{eq:boundsum_der3},
\begin{align*}
q_{0}''\brac{t} & =  \sum_{t_k \in T}\alpha_k \Ker^{\brac{2}}\brac{t-t_k} + \sum_{t_k \in T}\beta_k \Ker^{\brac{3}}\brac{t-t_k}\notag\\
  & \leq \brac{1+\normInfInf{\alpha- e_{t_j}}}K^{\brac{2}}\brac{t}\\
&  \quad+ \normInfInf{\alpha- e_{t_j}}\sum_{t_k \in T\setminus \{0\}}\abs{\Ker^{\brac{2}}\brac{t-t_k}} \\
&  \quad + \normInfInf{\beta}\Big(\abs{\Ker^{\brac{3}}\brac{t}}+  \sum_{t_k \in T\setminus \{0\}} \abs{\Ker^{\brac{3}}\brac{t-t_k}}\Big) \notag\\
& \leq - 2.30\, \flo^2. 
\end{align*}
Similar computations yield $ \abs{q_{0}''\brac{t}} \leq 4.07 \, \flo^2 $. This together with $q_{0}(0)=1$ and $q_{0}'(0)=0$ implies the desired result.
\end{proof}
\emph{\begin{lemma}
\label{lemma:qtau_3}
For any $t_l \in T \setminus \keys{t_j}$ and $t$ obeying $\abs{t-t_l} \leq
c \lambdaloEq$, we have 
\begin{align*}
|q_{t_j}(t)| & \leq 16.64 \brac{t-t_l}^2 \flo^2. 
\end{align*}
\end{lemma}}
\begin{proof}
We assume without loss of generality that $t_l = 0$ and prove the claim for $t \in (0,c \,\lambdalo]$. 
By \eqref{eq:boundAlpha}, \eqref{eq:boundBeta}, \eqref{eq:boundK}, \eqref{eq:boundsum_der2} and \eqref{eq:boundsum_der3}
\begin{align*}
  \abs{q_{t_j}''\brac{t}} & =  \abs{\sum_{t_k \in T}\alpha_k \Ker^{\brac{2}}\brac{t-t_k} + \sum_{t_k \in T}\beta_k \Ker^{\brac{3}}\brac{t-t_k}}\notag\\
& \leq \brac{1+\normInfInf{\alpha -  e_{t_j}}}\abs{K^{\brac{2}}\brac{t-t_j}}\\
& \quad + \normInfInf{\alpha- e_{t_j}} \brac{\abs{\Ker^{\brac{2}}\brac{t}} +  \sum_{t_k \in T\setminus \{0,t_j\}} \abs{\Ker^{\brac{2}}\brac{t-t_k}} } \\
& \quad + \normInfInf{\beta} \brac{\abs{\Ker^{\brac{3}}\brac{t}}+  \sum_{t_k \in T\setminus \{0\}} \abs{\Ker^{\brac{3}}\brac{t-t_k}}}\\
& \leq  16.64\, \flo^2, 
\end{align*}
since in the interval of interest $\abs{K^{(2)}\brac{t-t_j}} \leq \frac{18 \pi^2}{\fc^2 (\Deltamin-0.16\fc)^4} \leq 15.67 \,\fc^2$ due to Lemma~2.6 in~\cite{superres}. 
This together with $q_{t_j}(0)=0$ and $q_{t_j}'(0)=0$ implies the desired result.
\end{proof}

\section*{Acknowledgements}
This work was supported by a Fundaci\'{o}n Caja Madrid
Fellowship. The author is grateful to Emmanuel Cand\`es for useful comments regarding this manuscript and for his support.


\end{document}